\documentclass[a4paper,11pt]{article}
\pdfoutput=1
\usepackage{jheppub} 
\usepackage[english]{babel}
\usepackage{amsmath}
\usepackage{graphicx}
\usepackage{color}
\usepackage{mathrsfs}  
\usepackage[dvipsnames]{xcolor}
\usepackage{amssymb}
\usepackage{hyperref}
\usepackage{dcolumn}
\usepackage{bm}
\usepackage{mathtools,braket,blkarray}
\usepackage{multirow}
\usepackage{relsize}
\usepackage[normalem]{ulem}
\usepackage{tikz} 
\usetikzlibrary{arrows} 
\usetikzlibrary{decorations.markings} 

\tikzset{
	fermion/.style={postaction={decorate}, 
	   decoration={markings,mark=at position .575 with {\arrow{triangle 45}}}}, 
	 scalar/.style={dashed,postaction={decorate},
           decoration={markings,mark=at position .575 with {\arrow{triangle 45}}}}
}

\usepackage{soul}
\usepackage{subfig}
\usepackage{soul}

\newcommand {\be} {\begin{equation}}
\newcommand {\ee} {\end{equation}}

\newcommand{\GeV}{\,\mathrm{GeV}}
\newcommand{\eV}{\,\mathrm{eV}}

\definecolor{greenLinks}{rgb}{0, 0.6, 0} 
\definecolor{blueLinks}{rgb}{0, 0, 0.6}
\definecolor{redLinks}{rgb}{0.6, 0, 0}
\definecolor{tempText}{rgb}{0.55, 0.10,0.67}
\definecolor{eprintLinks}{rgb}{0.4, 0.4, 0.4}
\definecolor{journalLinks}{rgb}{0.6, 0, 0}
\definecolor{TitleBlue}{RGB}{56, 163, 165}

\def\vev#1{\left\langle #1\right\rangle}

\usepackage{float}

\def\vev#1{\left\langle #1\right\rangle}

\def\21{$\mathrm{SU(2)_L \otimes U(1)_Y}$ }
\def\31{$\mathrm{SU(3)_c \otimes U(1)_Q}$ }

\def\3211{$\mathrm{SU(3) \otimes SU(2)_L \otimes U(1)_R \otimes U(1)_{B-L}}$ }
\def\321{$\mathrm{SU(3) \otimes SU(2) \otimes U(1)}$ }
\def\422{$\mathrm{SU(4) \otimes SU(2) \otimes SU(2)_R}$ }
\newcommand {\ignore}[1]{}

\def\vev#1{\left\langle #1\right\rangle}

\def\U1{$\mathrm{ U(1)_{B_3 - 3 L_\mu} }$}

\newcommand{\AddrUNAM}{ Instituto de F\'{\i}sica, Universidad Nacional Aut\'onoma de M\'exico, A.P. 20-364, Ciudad de M\'exico 01000, M\'exico.}

\newcommand{\AddrAHEP}{%
 Institut de F\'{i}sica Corpuscular (CSIC-Universitat de Val\`{e}ncia), AHEP Group,\\ Parc Cient\'ific de Paterna.
 C/ Catedr\'atico Jos\'e Beltr\'an, 2 E-46980 Paterna (Valencia), Spain}
 \newcommand{\AddrMPIK}{%
  Max-Planck-Institut f\"{u}r Kernphysik, Saupfercheckweg 1, 69117 Heidelberg, Germany}

\newcommand{\AddrCatolica}{
  Departamento de Física, Universidad Católica del Norte,\\
Avenida Angamos 0610, Casilla 1280, Antofagasta, Chile}
\newcommand{\vone}{v_{\eta_1}}
\newcommand{\vones}{v^2_{\eta_1}}
\newcommand{\ves}{v_{s}}
\newcommand{\vess}{v^2_{s}}
\title{\color{TitleBlue}{Discrete dark matter mechanism as the source of neutrino mass scales}}

\author[a]{Cesar Bonilla,}
\author[b]{Johannes Herms,}
\author[c]{Omar Medina,}
\author[d]{Eduardo Peinado}

\affiliation[a]{\AddrCatolica}
\affiliation[b]{\AddrMPIK}
\affiliation[c]{\AddrAHEP}
\affiliation[d]{\AddrUNAM}

\emailAdd{cesar.bonilla@ucn.cl}
\emailAdd{herms@mpi-hd.mpg.de}
\emailAdd{Omar.Medina@ific.uv.es}
\emailAdd{epeinado@fisica.unam.mx}

\abstract{ The hierarchy in scale between atmospheric and solar neutrino mass splittings is investigated through two distinct neutrino mass mechanisms from tree-level and one-loop-level contributions. We demonstrate that the minimal discrete dark matter mechanism contains the ingredients for explaining this hierarchy. This scenario is characterized by adding new RH neutrinos and $SU(2)$-doublet scalars to the Standard Model as triplet representations of an $A_4$ flavor symmetry. The $A_4$ symmetry breaking, which occurs at the electroweak scale, leads to a residual $\mathbb{Z}_2$ symmetry responsible for the dark matter stability and dictates the neutrino phenomenology. Finally, we show that to reproduce the neutrino mixing angles correctly, it is necessary to violate CP in the scalar potential.
}

\begin{document}
\maketitle

\begin{abstract}

\end{abstract}


\section{Introduction}
Neutrino masses, the existence of dark matter (DM), and baryon asymmetry in the Universe are some of the most robust evidence for physics beyond the Standard Model (SM). Another puzzle that has long been attempted to be solved is the large mass hierarchy of charged fermions. The strongest hierarchical structure is shown by the up-type quarks, where at the $M_Z$ scale $m_t:m_c:m_u\sim 1:3.65\times 10^{-3}:7.5\times 10^{-6}$, while, in the down sector, $m_b:m_s:m_d\sim1:1.9\times 10^{-2}:9.64\times 10^{-5}$ and $m_\tau:m_\mu:m_e\sim 1:5.88\times 10^{-2}:2.79\times 10^{-4}$~\cite{Antusch:2013jca}. The most popular explanations are the Froggatt-Nielsen mechanism and the extra-dimensional approach of Randal-Sundrum~\cite{Froggatt:1978nt,Randall:1999ee}. In contrast, the masses in the neutrino sector are significantly less hierarchical. There have been some attempts to explain the hierarchy in the neutrino mass scales, which become manifest in two of the neutrino oscillation parameters, namely the atmospheric $\lvert \Delta m^{2}_{31} \rvert\sim 2.55\times 10^{-3}\eV^2$ (NO), and the solar mass splittings, $\Delta m^{2}_{21}\sim 7.5\times 10^{-5}\eV^2$, which differ by two orders of magnitude~\cite{deSalas:2020pgw,Esteban:2020cvm}. One possible explanation is that one of these mass hierarchies arises from a tree-level mass mechanism, while the other arises from a radiative mechanism, as in the scotogenic model~\cite{Ivanov:2017bdx,Rojas:2018wym,Aranda:2018lif}.
Another possible approach is supersymmetric models with R-parity violation, in which one neutrino receives mass at tree-level and the other two at one-loop~\cite{Hirsch:2000ef}.
In the type-I seesaw mechanism, the (light) left-handed neutrino mass matrix resulting from a single RH neutrino is rank-1. Hence it cannot explain the two squared mass differences ($\lvert \Delta m^{2}_{31} \rvert$, and $\Delta m^{2}_{21}$) unless it includes at least two RH neutrinos~\cite{Minkowski:1977sc,Schechter:1980gr}.
While the type-I seesaw mechanism may be economical; little can be learned, in its context, from the flavor information contained in the observed neutrino oscillation parameters.

Here, we discuss an extension of the SM with an $A_4$ flavor symmetry. The model has two beyond the standard model (BSM) fields, an $A_4$ triplet representation of RH neutrinos and an $A_4$ triplet scalar $SU(2)_L$-doublet. Only one RH neutrino is active when the flavor symmetry is spontaneously broken into a $\mathbb{Z}_2$ symmetry at the electroweak scale. At the same time, the other two are inert states that we will denominate ``dark'' fields.
We propose that the symmetry that explains the mixing angles of neutrinos is also responsible for the stability of dark matter in the context of the ``discrete dark matter'' mechanism (DDM) ~\cite{Hirsch:2010ru}.
In the original DDM, one of us, with collaborators, performed their neutrino mass analysis only at the tree level, which forced them to include an additional (fourth) RH neutrino to produce two massive neutrinos. In this work, we have gone a step further by analyzing all contributions to the light neutrino mass matrix at the one-loop level, taking into account the relevance they can potentially acquire~\cite{AristizabalSierra:2011mn,Grimus:2002nk}. This reveals two important facts.
The first is that assuming that the residual $\mathbb{Z}_2$ from the breaking of $A_4$ is conserved, a ``scotogenic'' mass mechanism~\cite{Ma:2006km} arises naturally.
Therefore the triplet of RH neutrinos can generate the two neutrino mass splittings, $\lvert \Delta m^{2}_{31} \rvert$ and $\Delta m^{2}_{21}$.
Second, the interplay between the seesaw and scotogenic mass mechanisms arises directly from the $A_4$ flavor symmetry in the discrete dark matter model. We argue that this interplay gives a natural explanation for the origin of the hierarchy between the two mass splittings ($\lvert \Delta m^{2}_{31} \rvert \gg \Delta m^{2}_{21}$) based on flavor symmetry.\\

\section{The Minimal Discrete Dark Matter Model}
We extend the SM symmetry group by an $A_4$ flavor symmetry, and the field content is minimally extended by three  $SU(2)_L$ doublet scalar fields in the triplet representation of $A_4$, $\eta = (\eta_1,\, \eta_2,\, \eta_3)$, and three RH neutrinos also as an $A_4$ triplet, $N_T = (N_1, N_2, N_3)$.
The quark sector is taken to be invariant under $A_4$ and only the $SU(2)_L$ Higgs doublet $H$ contributes to the quark masses. The fields associated with the leptons $e,\mu,\tau$ are assigned the three different $A_4$ singlet representations $1^{\prime\prime},\,1,\,1^\prime$ respectively, in such a way that the charged lepton mass matrix is diagonal. Exchanging the representation for the leptons results in similar predictions, see Appendix~\ref{Appendix:OtherScenarios}.
The relevant quantum numbers for the fields in the model are summarized in Table~\ref{tab:Mod}. In this section, we lay out the structure of Yukawa interactions constrained by the flavor symmetry, the spontaneous breakdown of $A_4$, and the scalar field spectrum.

\begin{table}[h!]
\begin{center}
\begin{tabular}{|c|c|c|c|c|c|c|c|c|c|c|}
\hline
& $\,L_e\,$ & $\,L_\mu \,$ & $\,L_\tau \,$ & $\,\,l_e\,\,$ & $\,\,l_\mu\,\,$ & $\,\,l_\tau\,\,$ & $N_T\,$ & $\,H\,$ & $\,\eta \,$ \\
\hline
$SU(2)$ & 2 & 2 & 2 & 1 & 1 & 1 & 1 &  2 & 2 \\
\hline
$U(1)_Y$ & $-\frac{1}{2}$ &  $-\frac{1}{2}$ &  $-\frac{1}{2}$ & -1 & -1 & -1 & 0 &   $\frac{1}{2}$ & $\frac{1}{2}$ \\
\hline
$A_4$ & $1^{\prime \prime} $ & $1$ & $1^{\prime}$ & $1^{\prime \prime}$ & $1$ & $1^{\prime}$ & 3 & 1 & 3  \\
\hline
\end{tabular}\caption{ Summary of the relevant particle content and quantum numbers of the model.} \label{tab:Mod}
\end{center}
\end{table}

\subsection{The Yukawa Sector}
The lepton sector Yukawa Lagrangian of the model is the following
\begin{equation}
\mathcal{L}^H_{\text{Yukawa}} =
y_e \overline{L}_e l_e H+y_{\mu} \overline{L}_{\mu} l_{\mu} H +y_{\tau} \overline{L}_{\tau} l_{\tau} H + \textit{h.c.},
\label{eq:YukH}
\end{equation}
\begin{equation}
\mathcal{L}^{\eta}_{\text{Yukawa}} =
 y_1^\nu \overline{L}_e (N_T\, \tilde{\eta})_{\bf{1}^{\prime \prime}} + y_2^\nu \overline{L}_\mu (N_T\,  \tilde{\eta}) _{\bf{1}} + y_3^\nu \overline{L}_\tau (N_T\,  \tilde{\eta})_{\bf{1}^{\prime}} + \frac{1}{2}m_{N}\overline{N^c_T} N_T+ \textit{h.c.},
 \label{eq:Yuketa}
\end{equation}
where $ \tilde{\eta}=i \tau_2 \eta^\dagger$, and the subscript in the parenthesis denotes the $A_4$ contraction of two triplets, see Appendix \ref{Appendix:A4} for details about the $A_4$ group, and the basis used for the generators. In Eq. (\ref{eq:Yuketa}), the Majorana mass term for the heavy right-handed (RH) neutrinos $N_T$  is included. Notice that due to the flavor symmetry, the three RH-neutrinos are degenerate, with mass 
 $m_N$.
 Since the charged leptons only couple to $H$, their mass matrix is proportional to its vacuum expectation value (vev), which we label by $v_H$. The $A_4$ symmetry forces the charged leptons' mass matrix to be diagonal
\begin{equation}
\label{eq:Mcharged}
 v_H Y^H_{l} = 
 v_H  \begin{pmatrix}
  y_e  & 0 & 0 \\
 0 & y_{\mu}  & 0  \\
  0 & 0  & y_{\tau} 
  \end{pmatrix}.
\end{equation}
From Eq. (\ref{eq:Yuketa}) the Yukawa coupling matrices of the $\eta$ fields with neutrinos are
\begin{equation}
  Y^{\eta_1}=\begin{pmatrix}
  y_1^\nu  & 0 & 0 \\
  y_2^\nu  & 0 & 0 \\
  y_3^\nu  & 0 & 0 
  \end{pmatrix}, \quad
   Y^{\eta_2}=\begin{pmatrix}
  0&y_1^\nu \omega^2 & 0 \\
  0&y_2^\nu  & 0 \\
  0&y_3^\nu \omega  & 0 
  \end{pmatrix}, \quad
  Y^{\eta_3}=\begin{pmatrix}
  0&0&y_1^\nu \omega   \\
  0&0&y_2^\nu  \\
  0&0&y_3^\nu \omega^2 
  \end{pmatrix}, \quad \text{where} \quad \omega=e^{i\frac{2 \pi}{3}}
\label{eq:YukEtaCouplings}
\end{equation}
their structure is dictated by the invariance of the Yukawa interactions under $A_4$.
We want to remark that there are only three parameters $y^{\nu}_1$, $y^{\nu}_2$, and $y^{\nu}_3$, ruling the Yukawa interaction of neutrinos with the three scalar doublets $\eta_1$, $\eta_2$, and $\eta_3$. 
This is a crucial feature of this model since, as detailed in the following sections, neutrino masses are generated by a type-I seesaw and a scotogenic mechanism; both share the same Yukawa couplings due to the flavor symmetry. Therefore, the dominance of the tree-level seesaw over the one-loop scotogenic mechanism emerges naturally from the $A_4$ invariance. This novel property can potentially explain the hierarchy between the neutrino mass splittings. In previous works that include a scoto-seesaw mechanism that is not based on non-Abelian flavor symmetries, the dominance between the mechanisms depends on the hierarchies between different Yukawa couplings \cite{Ibarra:2011gn,Rojas:2018wym,Aranda:2018lif,Barreiros:2020gxu,Barreiros:2022aqu,Mandal:2021yph,Ganguly:2022qxj}.

\subsection{Flavor Symmetry Breakdown}

The scalar potential of this model is written explicitly in Appendix~\ref{Appendix:scalarpot}. The $A_4$ symmetry is spontaneously broken at the electroweak scale by the vacuum expectation value of the $\eta$ fields\footnote{It is possible to break the flavor symmetry at a heavier scale by introducing flavon fields~\cite{Lamprea:2016egz,DeLaVega:2018bkp}.}. In the case that only $\eta_1$ acquires a vev, there is a residual $\mathbb{Z}_2$ symmetry that stabilizes a dark matter particle candidate~\cite{Hirsch:2010ru}
\begin{equation}
\vev{H^0} = \frac{v_H}{\sqrt{2}} \ne 0,\qquad
\vev{ \eta^0_1} = \frac{v_{\eta_1}}{\sqrt{2}} \ne 0,\qquad
\vev{\eta^0_{2,3}} = 0.
\label{eq:Vevs}
\end{equation} 
In the $A_4$ basis outlined in Appendix \ref{Appendix:A4}, the residual $\mathbb{Z}_2$ symmetry is manifest. The vev alignment in this model
\begin{equation}
\vev{\eta^0}=\frac{1}{\sqrt{2}}\begin{pmatrix} v_{\eta_1} \\ 0 \\0 \end{pmatrix},
\label{eq:etaVEV}
\end{equation}
remains invariant under the $S$ generator transformation 
\begin{equation}
S\left(\langle \eta^0 \rangle \right)= 
\begin{pmatrix}
1&0&0\\
0&-1&0\\
0&0&-1
\end{pmatrix}
\begin{pmatrix} \frac{v_{\eta_1}}{\sqrt{2}} \\ 0 \\0 \end{pmatrix}= \langle \eta^0 \rangle, \qquad S\left(\langle H^0 \rangle \right) = \langle H^0 \rangle .
\end{equation}
It is worth noting that the vev of the $A_4$ triplet $\eta$, as described in Eq. (\ref{eq:etaVEV}), belongs to the global minima of $A_4$ symmetric 3HDMs. This particular aspect has been thoroughly studied in the literature, as evidenced by these works \cite{Degee:2012sk,deMedeirosVarzielas:2022kbj,Carrolo:2022oyg}.

We define the ``dark'' fields, odd under this $\mathbb{Z}_2$, as some members of the $A_4$ triplets.
The $S$ generator dictates how the fields transform under the residual $\mathbb{Z}_2$ symmetry. In this way the ``dark'' fields are
\begin{align}
\label{eq:DarkFields}
\mathbb{Z}_2: \quad &\eta_2 \longrightarrow -\eta_2 \nonumber, \quad \hspace{6pt}\eta_3 \longrightarrow -\eta_3, \\
\qquad &N_2 \longrightarrow -N_2, \quad N_3 \longrightarrow -N_3,
\end{align}
while the rest of the fields are even, including the SM particles, and are referred to as ``active'' fields.

To analyze the scalar spectrum of the model, we can define the notation for the fields after electroweak symmetry breaking. Notice that the primed fields are written on the flavor basis
\begin{equation}\begin{array}{cc}
 H=\left(
\begin{array}{c}
H^{\prime +}_0\\
(v_H+H^{\prime}_0+i A^{\prime}_0)/\sqrt{2}
\end{array}
\right),
&\eta_1=\left(
\begin{array}{c}
H^{\prime +}_1\\
(v_{\eta_1}+H'_1+iA'_1)/\sqrt{2}
\end{array}
\right),\\ \\
\eta_2=\left(
\begin{array}{c}
H^{\prime +}_2\\
(H'_2+iA'_2)/\sqrt{2}
\end{array}
\right),&\eta_3=\left(
\begin{array}{c}
H^{\prime +}_3\\
(H'_3+iA'_3)/\sqrt{2}
\end{array}
\right).
\end{array}
\label{eq:ScalarDOF}
\end{equation}
The SM vev is given by $
  v^2={v^2_{\eta_1}}+{v^2_H} \simeq (246 \GeV)^2$,
following the customary notation, we define
\begin{equation}
\tan \beta = \frac{v_{H}}{v_{\eta_1}}.
\label{eq:tanbeta}
\end{equation}

One can perform a basis change among the fields to determine the mass-eigenstates. The mass matrix of the neutral scalars on the basis \begin{equation}
\{H_0^{\prime},\hspace{2pt}H_1^{\prime},\hspace{2pt}A_0^{\prime},\hspace{2pt}A_1^{\prime},\hspace{2pt}H_2^{\prime},\hspace{2pt}H_3^{\prime},\hspace{2pt}A_2^{\prime},\hspace{2pt}A_3^{\prime}\},
\label{eq:NeutralScalarBasis}
\end{equation}
has the following structure
\begin{equation}
M^{2}_{\text{neutral}}=\left(
\begin{array}{cccc}
M^2_{H_0^{\prime} H_1^{\prime}}&0&0&0\\
0&M^2_{A_0^{\prime}A_1^{\prime}}&0&0\\
0&0&M^2_{H_2^{\prime}H_3^{\prime}}&M^2_{\text{CPV}}\\
0&0&M^2_{\text{CPV}}&M^2_{A_2^{\prime}A_3^{\prime}}\\
\end{array}
\right).\quad 
\label{eq:blockdiag}
\end{equation}
The mass matrix of the charged scalars on the basis $\{H_0^{\prime+},\hspace{2pt}H_1^{\prime+},\hspace{2pt}H_2^{\prime+},\hspace{2pt}H_3^{\prime+}\}$ is 
\begin{eqnarray}\label{blockdiag}
M^2_{\text{charged}}\equiv\left(
\begin{array}{cc}
M^2_{H_0^{\prime+}H_1^{\prime+}}&0\\
0&M^2_{H_2^{\prime+}H_3^{\prime+}}\\
\end{array}
\right).\quad 
\end{eqnarray}
To have a clear notation, we define the mass eigenstate basis of these fields. Since there is no CP violation in the active sector, we have;
\begin{align}
\{H_0^{\prime},\hspace{2pt}H_1^{\prime}\} \quad  &\Longrightarrow \quad \{H_0, \hspace{2pt}h \}, 
\label{eq:BasisChange-Active-Neutral-Scalars}\\
\{A_0^{\prime},\hspace{2pt}A_1^{\prime}\} \quad  &\Longrightarrow \quad \{A_0, \hspace{2pt}G_A\}, \qquad \text{(mass eigenstates)}
\label{eq:BasisChange-Active-Neutral-PseudoScalars}\\
\{H_0^{\prime+},\hspace{2pt}H_1^{\prime+}\} \quad  &\Longrightarrow \quad \{H^+, \hspace{2pt}G^{+}_0 \}.
\label{eq:BasisChange-Active-Charged}
\end{align}
In the dark sector, there are in general CP-violating terms in the scalar potential. Therefore, neutral scalars and pseudoscalars generically mix.  We define the neutral field basis as follows
\begin{equation}
\{H_2^{\prime},\hspace{2pt}H_3^{\prime},\hspace{2pt}A_2^{\prime},\hspace{2pt}A_3^{\prime}\} \quad  \Longrightarrow \quad \{\chi^{D}_1, \hspace{2pt}\chi^D_2, \hspace{2pt}\chi^D_3,\hspace{2pt}\chi^D_4\} \quad \text{such that} \quad m_{\chi^{D}_1} \geq m_{\chi^{D}_2} \geq m_{\chi^{D}_3} \geq m_{\chi^{D}_4}.
\label{eq:BasisChange-Dark-Neutral}
\end{equation}
The lightest of these fields, $\chi^{D}_4$, is a dark matter candidate, stabilized by the $\mathbb{Z}_2$ symmetry (in the scenario $M_N \gg m_{\chi^{D}_4}$, which is the only viable case we found for this model).
Lastly, we define the mass eigenstates of the charged dark scalars
\begin{equation}
\{H_2^{\prime+},\hspace{2pt}H_3^{\prime+}\} \quad  \Longrightarrow \quad \{\chi^{D+}_1, \hspace{2pt}\chi^{D+}_2\} \quad \text{such that} \quad m_{\chi^{D+}_1} \geq m_{\chi^{D+}_2}. \label{eq:BasisChange-Dark-Charged}
\end{equation}

The expressions for the scalar mass matrices are written explicitly in Appendix~\ref{Appendix:scalarpot} in terms of the parameters of the scalar potential. The matrices $M^2_{A_0^{\prime} A_1^{\prime}}$ and $M^2_{H_0^{\prime+}H_1^{\prime+}}$ have a vanishing eigenvalue corresponding to the neutral ($G_A$) and charged ($G^+$) Goldstone bosons, respectively.

The diagonalization of the mass matrices in the scalar sector is very important for neutrino mass generation. The diagonalizing orthogonal or unitary matrices are defined as follows:
\begin{align}
D^2_{H_0h}=O_{H_0h} M^2_{H_0^{\prime} H_1^{\prime}} O^T_{H_0h}&;~~~~D^2_{A_0G}=O_{A_0G}M^2_{A_0^{\prime}A_1^{\prime}}O^T_{A_0G};\nonumber \\
D^2_{H_0^+G^+}=O_{H_0^+G^+}M^2_{H_0^{\prime+}H_1^{\prime+}}O^T_{H_0^+G^+}&;~~~~ D^2_{\chi^{D+}}=U_{\chi^{D+}}M^2_{H_2^{\prime+}H_3^{\prime+}}U^{\dagger}_{\chi^{D+}}; \label{eq:RotMat}\\
D^2_{\chi}= O_{\chi} &\begin{pmatrix}M^2_{H_2^{\prime}H_3^{\prime}}M^2_{\text{CPV}}\\M^2_{\text{CPV}} M^2_{A_2^{\prime}A_3^{\prime}}\end{pmatrix} O^T_{\chi}, \nonumber
\end{align}
with
\begin{equation}
O_{A_0 G}=O_{H_0^+G^+}=
\begin{pmatrix}
\cos\beta&-\sin\beta\\
\sin\beta&\cos\beta\\
\end{pmatrix}.
\end{equation}
The angle $\tan \beta$, defined in Eq. (\ref{eq:tanbeta}), parameterizes the mixing within the $\mathbb{Z}_2$-even sector for the charged scalars, and also for the pseudoscalars. 
We consider the scenario where the lighter of the active neutral scalars, $h$, is the Higgs boson observed at the LHC, given by
\begin{equation}
    \begin{pmatrix}
        H_0 \\ h
    \end{pmatrix}
    =
    O_{H_0 h}
    \begin{pmatrix}
        H_0' \\ H_1'
    \end{pmatrix}
    =
    \begin{pmatrix}
        \sin \alpha & \cos \alpha \\
        \cos\alpha & -\sin\alpha
    \end{pmatrix}
    \begin{pmatrix}
        H_0' \\ H_1'
    \end{pmatrix}
    \,,
\end{equation}
where $\alpha$ parameterizes the misalignment between $H_1'$ and the heavy neutral scalar $H$, in analogy to the type-I 2HDM.

\section{Neutrino Masses}
\label{Section:NeutrinoMasses}
In this model, both the active and dark sectors contribute to generate neutrino masses:
\begin{equation}
(m_{\nu})_{\alpha \beta}=(m^{\text{Active}}_{\nu})_{\alpha \beta}+(m^{\text{Dark}}_{\nu})_{\alpha \beta}
\label{eq:TotalNeutrinoMassMatrix}
\end{equation}
In this equation, the fields $\eta_1,~ H,~ N_1$ and  $Z$ will contribute to $m^{\text{Active}}_{\nu}$, while the fields in Eq. (\ref{eq:DarkFields}) contribute to $m^{\text{Dark}}_{\nu}$.
From this point on, we define the following set of indices, which will be used in this section 
\begin{equation}
\alpha,\beta=e,\mu,\tau \hspace{2pt}, \quad \text{and} \quad n=N_2,N_3, 
\label{eq:LeptonIndices}
\end{equation}
meaning that $n$ runs only through the dark RH neutrinos.

We analyze both active and dark contributions separately.
Our notation for the one-loop contributions, although slightly different, is based on an article in which the authors write the expression for the light neutrino mass matrix for a generalized scotogenic model with an arbitrary number of dark isodoublets and neutral heavy fermions 
 \cite{Escribano:2020iqq}, where, however, they assume a CP-symmetry-preserving scalar sector ab initio.
 This is in contrast to our model, where, as detailed below, CP-breaking in the scalar sector is required to fit the current experimental values of the lepton mixing parameters.

\subsection{Active Fields Contribution}
\begin{figure}    \centering
\includegraphics[width=3.5in]{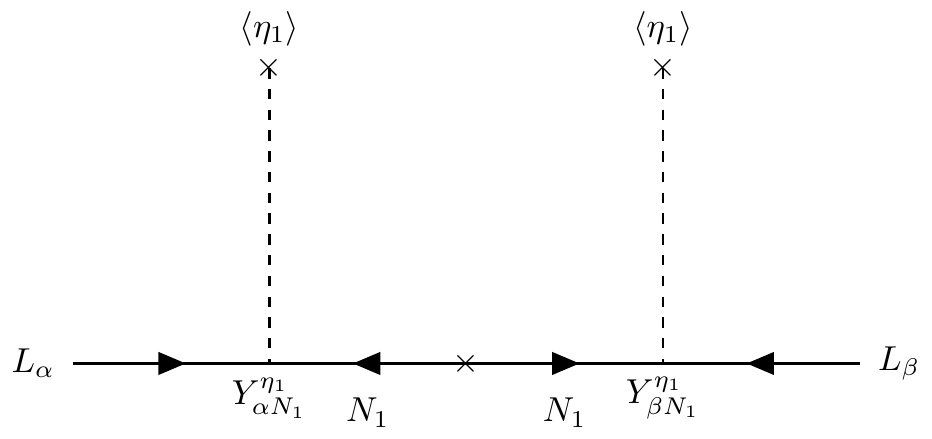}
    \caption{
    Feynman diagram of the tree-level seesaw. Only the active fields $N_1$ and $\eta_1$ partake in this mass mechanism, the Yukawa couplings $Y^{\eta_1}_{\alpha N_1}$ are written explicitly in Eq. (\ref{eq:YukEtaCouplings}). The contribution from this diagram can only generate a rank-1 mass matrix for light neutrinos. Together with its one-loop corrections in Figure \ref{fig:Active-One-Loop} generate the atmospheric mass splitting $\lvert \Delta m^{2}_{31} \rvert$ (NO).}
    \label{fig:ActiveSeesaw}
\end{figure}

From Eqs. (\ref{eq:Yuketa}), (\ref{eq:YukEtaCouplings}), and (\ref{eq:Vevs}) the Dirac neutrino mass matrix $m_D$, and the heavy Majorana neutrino mass matrix $M_{\text{R}}$ are given respectively by
\begin{equation}
\label{eq:MDirA}
 m_{D} = 
  \frac{\vone}{\sqrt{2}}\begin{pmatrix}
  y_1^\nu   & 0 & 0 \\
  y_2^\nu   & 0 & 0 \\
  y_3^\nu   & 0 & 0 
  \end{pmatrix}, \qquad 
   M_{\text{R}} = 
  \begin{pmatrix}
  m_N & 0 & 0 \\
  0 & m_N & 0 \\
  0 &0 & m_N 
  \end{pmatrix}.
\end{equation}
The three heavy Majorana neutrinos $N_T = (N_1, N_2, N_3)$ are degenerate. Only one heavy Majorana neutrino, $N_1$, participates in the seesaw mechanism (Figure~\ref{fig:ActiveSeesaw}). Its contribution to the neutrino masses is
\begin{equation}
m^{\text{Tree}}_{\nu} \approx - m_D {M_R}^{-1} {m_D}^{T},
\label{eq:mnuTree}
\end{equation}
this generates the mass matrix for light neutrinos 
\begin{equation} 
m^{\text{Tree}}_\nu
 = -\frac{v^2_{\eta_{1}}}{2 M }
\begin{pmatrix}
 y^{\nu}_1 y^{\nu}_1 & y^{\nu}_1 y^{\nu}_2  &  y^{\nu}_1 y^{\nu}_3  \\
 y^{\nu}_1 y^{\nu}_2 & y^{\nu}_2 y^{\nu}_2  &  y^{\nu}_2 y^{\nu}_3  \\
 y^{\nu}_1 y^{\nu}_3 & y^{\nu}_2 y^{\nu}_3  &  y^{\nu}_3 y^{\nu}_3  
  \end{pmatrix},
\label{eq:MnuATree}
\end{equation}
which is rank-1, meaning that only one SM neutrino is massive at tree-level.

\begin{figure}[]%
    \centering
    \subfloat[\centering One-loop correction from the active fields to the seesaw mechanism]{{ 
        \includegraphics[width=2.6in]{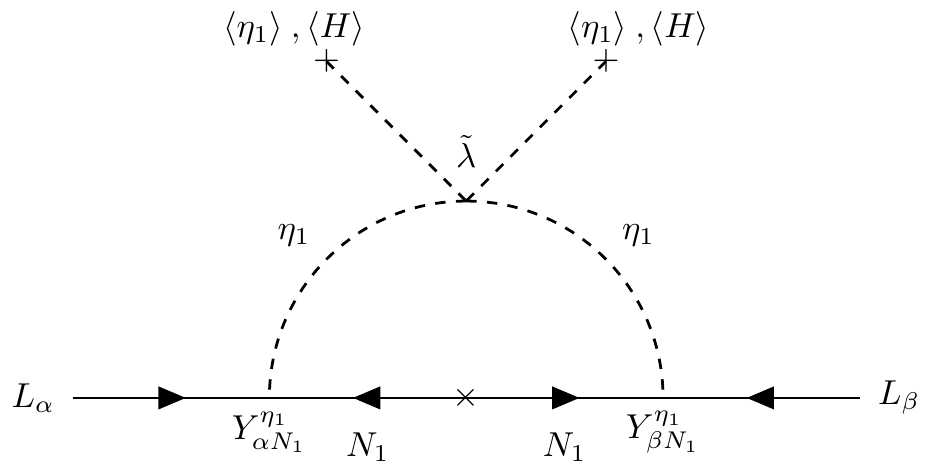}
}}%
    \subfloat[\centering 
    One-loop correction from the Z-boson to the seesaw mechanism]{{
     \includegraphics[width=2.6in]{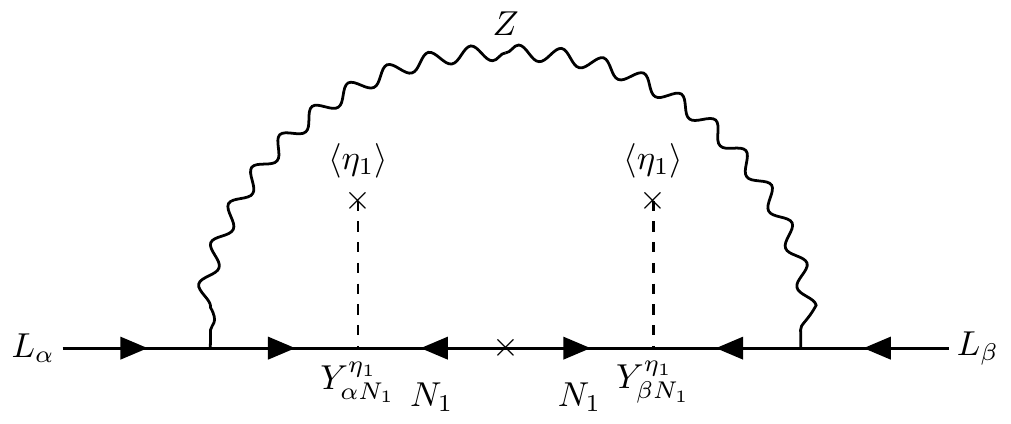}
}}%
    \caption{
    Corrections from the active fields $\eta_1$, $N_1$, and the Z-boson. Their contribution to the neutrino mass matrix is proportional to the tree-level seesaw mechanism. The $\tilde{\lambda}$ represents the relevant scalar potential couplings.}%
    \label{fig:Active-One-Loop}%
\end{figure}

The matrix $m^{\text{Active}}_{\nu}$ contains the tree-level contribution ($m^{\text{Tree}}_{\nu}$), as well as one-loop corrections from the different active fields (all involving $N_1$). We can write it as follows
\begin{equation}
m^{\text{Active}}_{\nu}= m^{\text{Tree}}_{\nu} + m^{\text{One-loop}}_{\nu, N_1} + m^{\text{One-loop}}_{\nu, Z}.
\label{eq:mnu-Active}
\end{equation}
The one-loop corrections to the seesaw mechanism from the active sector are shown in Fig.  \ref{fig:Active-One-Loop}.
It is important to notice that the matrices $m^{\text{One-loop}}_{\nu, N_1}$, and  $m^{\text{One-loop}}_{\nu, Z}$ are proportional to the tree-level seesaw mass matrix $m^{\text{Tree}}_{\nu}$ in Eq. (\ref{eq:MnuATree}), therefore they do not increase the rank of $m^{\text{Active}}_{\nu}$.

 From the Yukawa Lagrangians of the model, Eqs.  (\ref{eq:YukH}), and (\ref{eq:Yuketa}), we obtain the Yukawa coupling matrices for the different scalar degrees of freedom on the flavor basis, as defined in Eq. (\ref{eq:ScalarDOF}).
 For the active scalars  $\{H_0^{\prime},\hspace{2pt}H_1^{\prime},\hspace{2pt}A_0^{\prime},\hspace{2pt}A_1^{\prime}\}$, the only non-vanishing coupling matrices are those for the $N_1$ fermion with the scalar $H^{\prime}_1$ and the pseudoscalar $A^{\prime}_1$ fields, that is
\begin{equation}
Y^{H_1^{\prime}}=Y^{\eta_1}, \quad \mbox{and}\quad Y^{A_1^{\prime}}=iY^{\eta_1},
\label{eq:Yukawa-Active}
\end{equation}
 respectively. With this, we can write
\begin{equation}
(m^{\text{One-loop}}_{\nu, N_1})_{\alpha \beta} = -\frac{1}{32 \pi^2} \sum_{a} m_N Y^{a}_{\alpha N_1} Y^{a}_{\beta {N_1}} B_0(0,m^2_a,m_N),  \qquad a=h,H_0,G,A_0, 
\label{eq:mnuActive-main}
\end{equation}
where we use the Passarino-Veltman loop function $B_0$ \cite{Passarino:1978jh,Escribano:2020iqq};
\begin{equation}
B_{0}(0,m^2_a,m^2_{N}) = \Delta_{\epsilon} + 1 - \frac{m^2_a \log{m^2_a} - m^2_N \log{m^2_N} }{m^2_a - m^2_N},
\label{eq:B0}
\end{equation}
in which $\Delta_{\epsilon}$ diverges in the limit $\epsilon \longrightarrow 0$. And the only non-vanishing one-loop correction involving the Z-boson \cite{AristizabalSierra:2011mn}
\begin{equation}
(m^{\text{One-loop}}_{\nu, Z})_{\alpha \beta} = \frac{3}{16 \pi^2} \frac{m^2_{Z}}{v^2} \log \left(\frac{m^2_{Z}}{m^2_N}\right) (m^{\text{Tree}}_{\nu})_{\alpha \beta}.
\label{eq:mnuActive-Zboson}
\end{equation}
Note that the Yukawa coupling matrices in Eq. (\ref{eq:mnuActive-main}) are written on the scalar mass-eigenstate basis. These are obtained using the basis-change matrices defined in Eq. (\ref{eq:RotMat}). We have
\begin{equation}
Y^{a}_{\alpha n}= \left(O_{ H_0 h}\right)^{a}_{\hspace{3pt}k}Y^{k}_{\alpha n}, \qquad \text{for} \quad a= H_0, h \quad \text{and} \quad k=H^{\prime}_1,
\label{eq:YukActiveScalars}
\end{equation}
and
\begin{equation}
Y^{a}_{\alpha n}= \left(O_{A_0 G}\right)^{a}_{\hspace{3pt}k}Y^{k}_{\alpha n}, \qquad \text{for} \quad a= A_0, G \quad \text{and} \quad k= A^{\prime}_1.
\label{eq:YukActiveScalars}
\end{equation}

All the active field contributions add up to a rank-1 mass matrix and therefore generate only one massive neutrino. On the other hand, the scotogenic mass mechanism involving the dark fields generates the two lighter neutrino masses.

\subsection{Dark Fields Contribution}
\begin{figure}
    \centering    \includegraphics[width=3.5in]{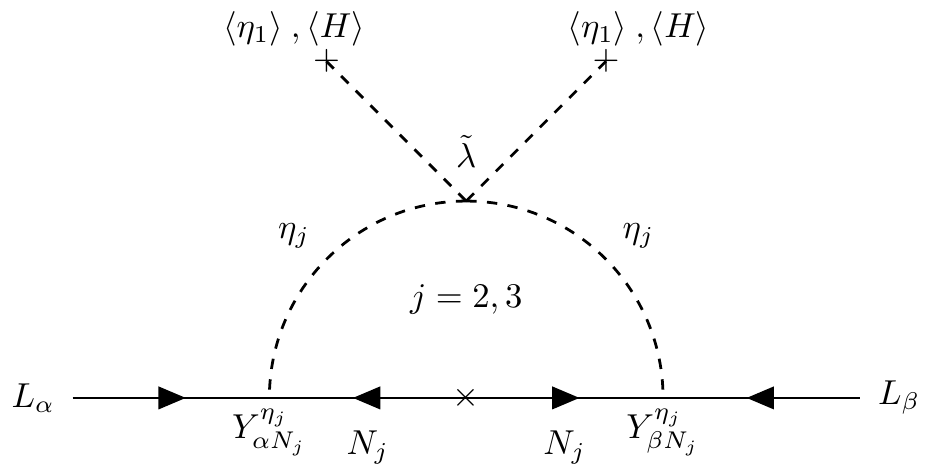}
    \caption{
    Feynman diagram of the scotogenic mechanism. The dark fermions $N_2$, and $N_3$; and the scalars $\eta_2$ and $\eta_3$ partake in this mass mechanism, the Yukawa couplings $Y^{\eta_j}_{\alpha N_j}$ for $j= 2,3$ are written explicitly in Eq. (\ref{eq:YukEtaCouplings}). The contribution from this diagram generates a rank-2 mass matrix for light neutrinos. This contribution generates radiatively the solar mass splitting $\Delta m^{2}_{21}$ (NO).
    }
    \label{fig:DarkScotogenic}
\end{figure}
The dark fields contribute to the neutrino mass matrix through the scotogenic mechanism~\cite{Ma:2006km} at one-loop level, as shown in Figure~\ref{fig:DarkScotogenic}.
This contribution is denoted by
\begin{equation}
m^{\text{Dark}}_{\nu}= m^{\text{One-loop}}_{\nu, N_n}.
\label{eq:mnu-Dark}
\end{equation}
The Yukawa coupling matrices for the dark scalars $\{ H^{\prime}_2, H^{\prime}_3, A^{\prime}_2,A^{\prime}_3 \}$ in the flavor basis (\ref{eq:YukEtaCouplings}) are given by
\begin{equation}
Y^{H^{\prime}_2}=Y^{\eta_2}, \quad Y^{H^{\prime}_3}=Y^{\eta_3}, \quad Y^{A^{\prime}_2}=iY^{\eta_2}, \quad Y^{A^{\prime}_3}=iY^{\eta_3}.
\label{eq:Yukawa-Dark}
\end{equation}
From this, the contribution to the neutrino mass matrix from the scotogenic mechanism is given by
\begin{equation}
(m^{\text{Dark}}_{\nu})_{\alpha \beta} = -\frac{1}{32 \pi^2} \sum_{n,a} m_N Y^{a}_{\alpha n} Y^{a}_{ \beta n} B_0(0,m^2_a,m^2_N), \qquad a=\chi^{D}_1,...,\chi^{D}_4,
\label{eq:mnuDark-main}
\end{equation}
with $B_0$ as written in Eq. (\ref{eq:B0}), and the Yukawa couplings in the mass-eigenstate basis
\begin{equation}
Y^{a}_{n\alpha}= \left(O_{\chi}\right)^{a}_{\hspace{3pt}k}Y^{k}_{n\alpha}, \qquad \text{for} \quad a=\chi^{D}_1,...,\chi^{D}_4, \quad \text{and} \quad k=H^{\prime}_2, H^{\prime}_3, A^{\prime}_2,A^{\prime}_3.
\label{eq:YukDarkScalars}
\end{equation}
According to the definition in Eq. (\ref{eq:BasisChange-Dark-Neutral}), the dark scalar $\chi^D_4$ is a dark matter candidate involved in the generation of neutrino masses, which is the essence of the scotogenic mechanism.
The dark mass matrix $m^{\text{Dark}}_{\nu}$ is rank-2, which naturally explains normal mass ordering (NO) scenario, with $m_3$ generated at tree-level, while $m_{1,2}$ are loop suppressed.\footnote{Neutrino oscillations and cosmology favor normal order~\cite{DeSalas:2018rby}}. We will comment further on this in the next section.

\section{Results}
We performed a thorough scan of the model's parameter space to investigate the suitability of the model describing the experimental data on measured quantities related to the flavor in the lepton sector.

In the most general scenario, where CP-symmetry is generically broken in the scalar sector, the model has the following set of real independent parameters
\begin{equation}
\{y_e,\hspace{2pt}y_{\mu},\hspace{2pt}y_{\tau},\hspace{2pt}y^{\nu}_{1},\hspace{2pt}y^{\nu}_{2},\hspace{2pt}y^{\nu}_{3},\hspace{2pt}v_{\eta_{1}},\hspace{2pt}v_{H},\hspace{2pt}m_{N},\hspace{2pt}\lambda_{1},\hspace{2pt}\lambda_{2},\hspace{2pt}\lambda_{3},\hspace{2pt}\lambda_{4},\hspace{2pt}\lambda_{5},\hspace{2pt}\lambda_{6},\hspace{2pt}\lambda_{7},\hspace{2pt}\lambda_{8},\hspace{2pt}\lambda_{9},\hspace{2pt}\lambda_{10},\hspace{2pt}\varphi_{5},\hspace{2pt}\varphi_{9},\hspace{2pt}\varphi_{10}\},
\label{eq:ModelParameters}
\end{equation}
which include the Yukawa couplings, which are real up to unphysical phases due to flavor symmetry, and the parameters and phases of the scalar potential (see Appendix \ref{Appendix:scalarpot}). The phases $\varphi_5$, $\varphi_9$, and $\varphi_{10}$ in last equation are the only source of CP violation in the model.

We analyzed the value of different observables obtained in terms of the parameters in Eq. (\ref{eq:ModelParameters}). We considered eleven measured observables $\mu_{exp}$, including the six neutrino oscillation parameters, the masses of the charged leptons, and the vev and mass of the Higgs field of the Standard Model 
\begin{equation}
\mu_{exp} \in \{ \sin^{2}\theta_{12},\hspace{2pt}\sin^{2}\theta_{13},\hspace{2pt}\sin^{2}\theta_{23},\hspace{2pt}\delta^{\ell},\hspace{2pt}\Delta m^2_{21},\hspace{2pt}\Delta m^2_{13},\hspace{2pt}m_e,\hspace{2pt}m_{\mu},\hspace{2pt}m_{\tau},\hspace{2pt}v,\hspace{2pt}m_h \}.
    \label{eq:muObs}
\end{equation}
 Lepton mixing experimental values are taken from the Neutrino Global Fit \cite{deSalas:2020pgw}, while charged lepton masses are taken at the Z-boson mass $m_{Z}$ scale \cite{Antusch:2013jca,ParticleDataGroup:2022pth}.
 Due to the large number of free parameters in the scalar potential, we do not obtain clear predictions for scalar sector observables. For the parameter scan, we demand perturbativity, positivity of the Hessian, the observed mass of the SM-like Higgs~\cite{ParticleDataGroup:2022pth}, and a mass $>300 \GeV$ for all non-DM BSM scalars.
 To ensure that the scalar potential is bounded from below~\cite{Boucenna:2011tj}, we use the parameters from Appendix~ \ref{Appendix:scalarpot}.
We provide a benchmark in Section~\ref{sec:benchmark}, for which we also discuss dark matter and Higgs constraints.

In the following subsections, we highlight some correlations between the lepton mixing parameters in Eq. (\ref{eq:muObs}), the lightest neutrino mass $m^{\nu}_{\text{lightest}}$, and the neutrinoless double beta decay mass parameter $\langle m_{\beta\beta} \rangle$. It's important to note that these are the only clear correlations we were able to identify between these observable quantities in our analysis.

\subsection{Normal Ordering of Neutrino Masses}
The model predicts normal ordering (NO) for neutrino masses since viable results for inverted order (IO) are not obtainable. This is a consequence of the interplay between the seesaw mechanism involving the active fields and the radiative scotogenic mass mechanism involving the dark fields.
The type-I seesaw mechanism appears at tree level (see Figures \ref{fig:ActiveSeesaw} and \ref{fig:Active-One-Loop}), and due to the $A_4$ symmetry can only generate one neutrino mass, since $m^{\text{Active}}_{\nu}$ in Eq.~(\ref{eq:mnu-Active}) is a rank-1 matrix. It is the main contribution to the heaviest neutrino mass $m^{\nu}_3$ and consequently to the atmospheric mass splitting
\begin{equation}
m^{\text{Active}}_{\nu} \quad \Longrightarrow \quad \Delta m^{2}_{31} \quad \text{(NO)}.
\label{eq:}
\end{equation}
The scotogenic mechanism (see Figure \ref{fig:DarkScotogenic}) generates $m^{\text{Dark}}_{\nu}$ in Eq.~(\ref{eq:mnu-Dark}), a rank-2 mass matrix that contributes to the two lightest neutrino masses $m^{\nu}_2$ and $m^{\nu}_1 =m^{\nu}_{\text{lightest}}$ and therefore to the solar mass splitting
\begin{equation}
m^{\text{Dark}}_{\nu} \quad \Longrightarrow \quad \Delta m^{2}_{21} \quad \text{(NO)}.
\label{eq:}
\end{equation}
In this model, the dominance of the tree-level seesaw mechanism over the scotogenic model is a prediction. This is not a generic feature from scoto-seesaw models \cite{Ibarra:2011gn,Rojas:2018wym,Aranda:2018lif,Barreiros:2020gxu,Barreiros:2022aqu,Mandal:2021yph,Ganguly:2022qxj}, it is a result from the fact that the mass mediators for both mechanisms in this model belong to the same multiplets of $A_4$, meaning that the Yukawa coupling matrices entering $m^{\text{Active}}_{\nu}$, and $m^{\text{Dark}}_{\nu}$ are the same parameters $y^{\nu}_1$, $y^{\nu}_2$ and $y^{\nu}_3$.

\subsection{Lepton CP-phase $\delta^{\ell}$}
In this model, there are only six Yukawa couplings that enter the charged lepton and neutrino mass matrices as described by Eqs. (\ref{eq:Mcharged}), and (\ref{eq:YukEtaCouplings}) respectively. These are in general complex parameters
\begin{equation}
\{ y_{e},\hspace{2pt}y_{\mu},\hspace{2pt}y_{\tau},\hspace{2pt}y^{\nu}_{1},\hspace{2pt}y^{\nu}_{2},\hspace{2pt}y^{\nu}_{3} \},
    \label{eq:YukawaCouplings}
\end{equation}
nevertheless, their associated phases are non-physical and can be reabsorbed by field re-definitions. Thus in this model lepton CP-violation has two properties:
the first one is that lepton CP-symmetry is strictly preserved at the tree-level and therefore can only emerge from the scalar sector phases (see Appendix \ref{Appendix:scalarpot})
\begin{equation}
\{\varphi_5,\hspace{2pt}\varphi_9,\hspace{2pt}\varphi_{10}\},
\label{eq:ScalarPotPhases}
\end{equation}
through radiative corrections.
The second one is that from our analysis of the parameter space, we found that assuming CP conservation in the scalar potential Eq.~(\ref{eq:ScalarPotential}) (ie.~setting the phases in Eq. (\ref{eq:ScalarPotPhases}) to zero) is ruled out.
In this CP-conserving scenario the mixing between the dark scalars $\eta_2$, $\eta_3$ is fixed and the model is unable to reproduce the oscillation parameters in Eq.~(\ref{eq:muObs}), i.e., this model requires CP violation in the scalar sector to be viable. To illustrate this, we summarize the best fit point for this 
case Appendix~\ref{Appendix:OtherScenarios}.

Nonetheless, the results from scanning the parameter space
give values the lepton CP-phase $\delta^{\ell}$ close to the lepton CP-conservation
\begin{equation}
\delta^{\ell} \sim 0,\hspace{2pt} 2 \pi, \qquad \delta^{\ell} \sim \pi\,.
    \label{eq:CPPreserv}
\end{equation}\\

\begin{figure}
    \centering    \includegraphics[width=3.5in]{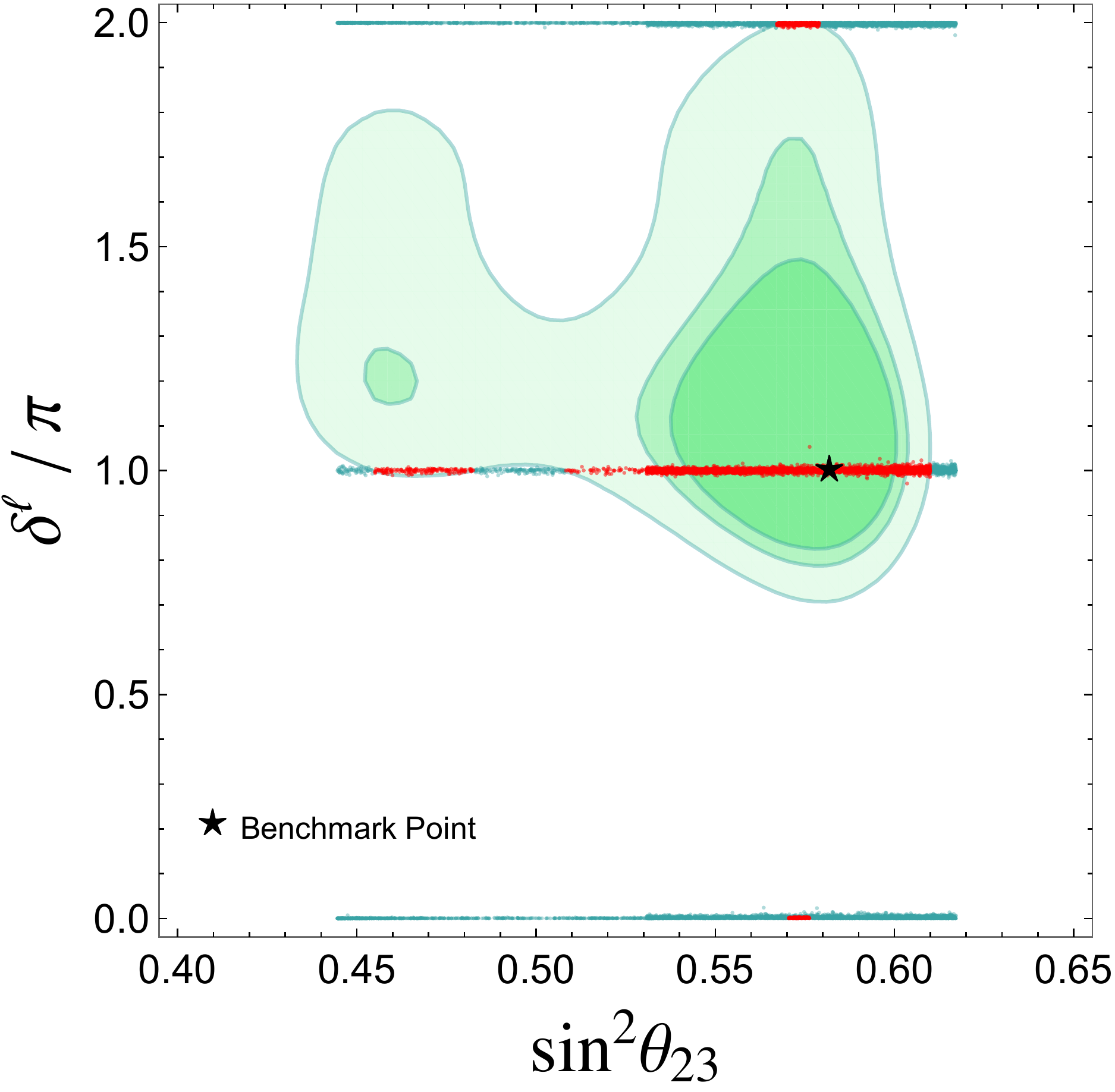}
    \caption{
    All the points displayed are in the parameter space of the model and inside the $3 \sigma$ range of all the observables in Eq. (\ref{eq:muObs}). The regions in different shades of green  are the $90$, $95$, and $99\%$ CL regions in the $\sin^{2}\theta_{23}$ vs. $\delta^{\ell}$ plane. Table \ref{tab:fit} displays the benchmark point data. The experimental values are taken from the Neutrino Global Fit \cite{deSalas:2020pgw}.
    }
\label{fig:AtmVsCP}
\end{figure}

This can be seen in Figure \ref{fig:AtmVsCP}, where all the plotted points are inside the $3\sigma$ region of all the observables listed in Eq.~(\ref{eq:muObs}). Furthermore, the different shades of green represent $90$, $95$, and $99\%$ CL regions in the $\sin^{2}\theta_{23}$ vs. $\delta^{\ell}$ plane, the points in red are inside this region. 
This color convention for the points will appear on all the other scatter plots in this work
\cite{deSalas:2020pgw}. 

\subsection{Lightest Neutrino Mass, and $\nu0\beta\beta$ Decay}
\begin{figure}[hbt]
    \centering    \includegraphics[width=3.5in]{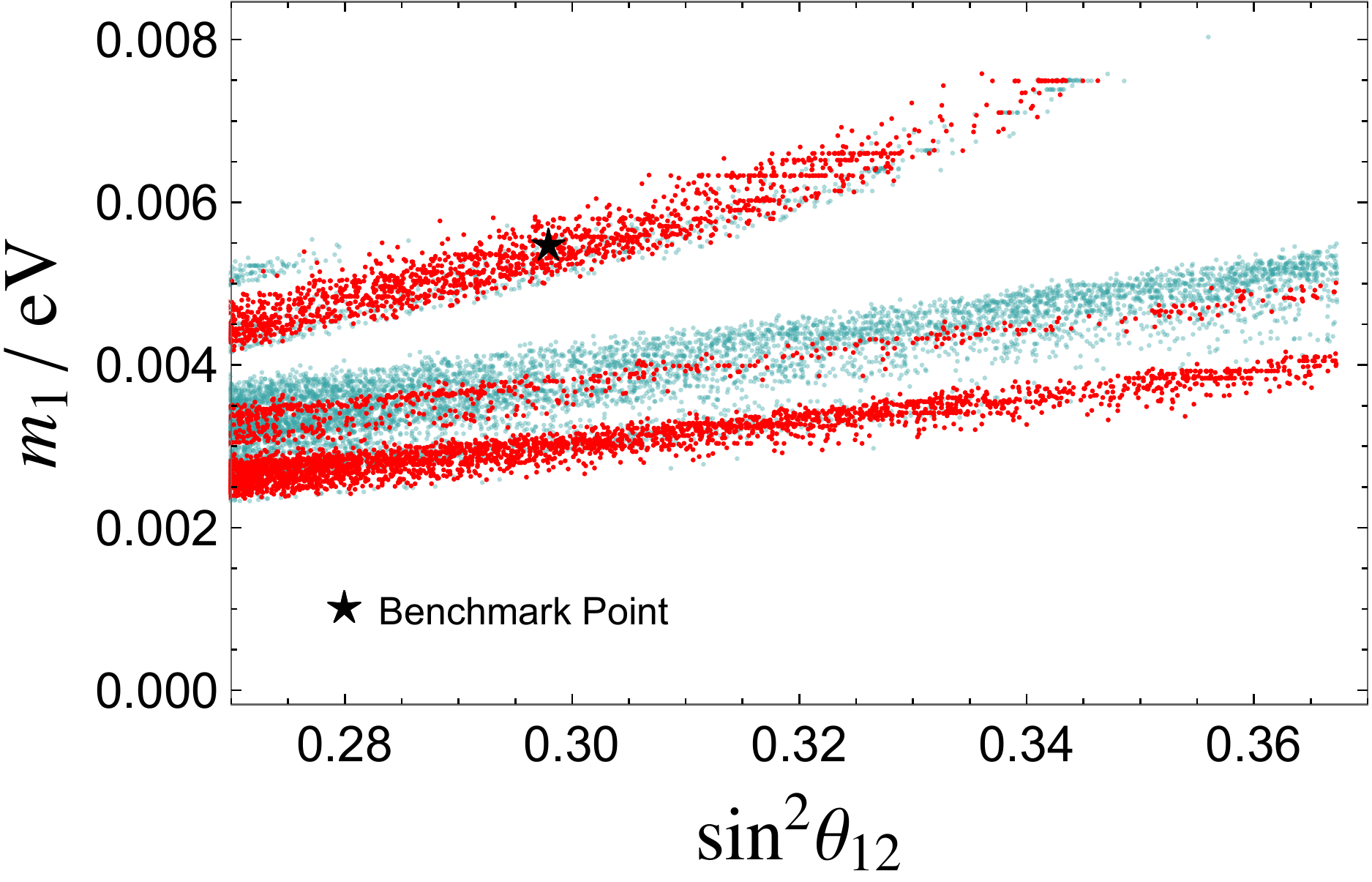}
    \caption{
    There is a strong correlation between the mass of the lightest neutrino $m^{\nu}_1$ and the solar mixing angle $\sin ^{2}\theta_{12}$ in this model. This correlation constrains the $m^{\nu}_1$ attainable range. The coloring of the points matches Figure \ref{fig:AtmVsCP}.
    }
\label{fig:SolVsm1}
\end{figure}

We obtain that the value for the lightest neutrino mass $m^{\nu}_1 = m^{\nu}_{\text{lightest}}$ (NO) is constrained in this model to the range
\begin{equation}
 2\hspace{2pt} \text{meV}\hspace{2pt}\lesssim \hspace{2pt} m^{\nu}_{\text{lightest}} \hspace{2pt} \lesssim \hspace{2pt} 8 \hspace{2pt} \text{meV}.
\label{eq:NuMassScale}
\end{equation}
This is a consequence of a strong correlation between the lightest neutrino mass $m^{\nu}_1$ and the solar mixing angle $\sin^{2} \theta_{12}$, as displayed in Figure \ref{fig:SolVsm1}. 

\begin{figure}[hbt]
    \centering    \includegraphics[width=3.5in]{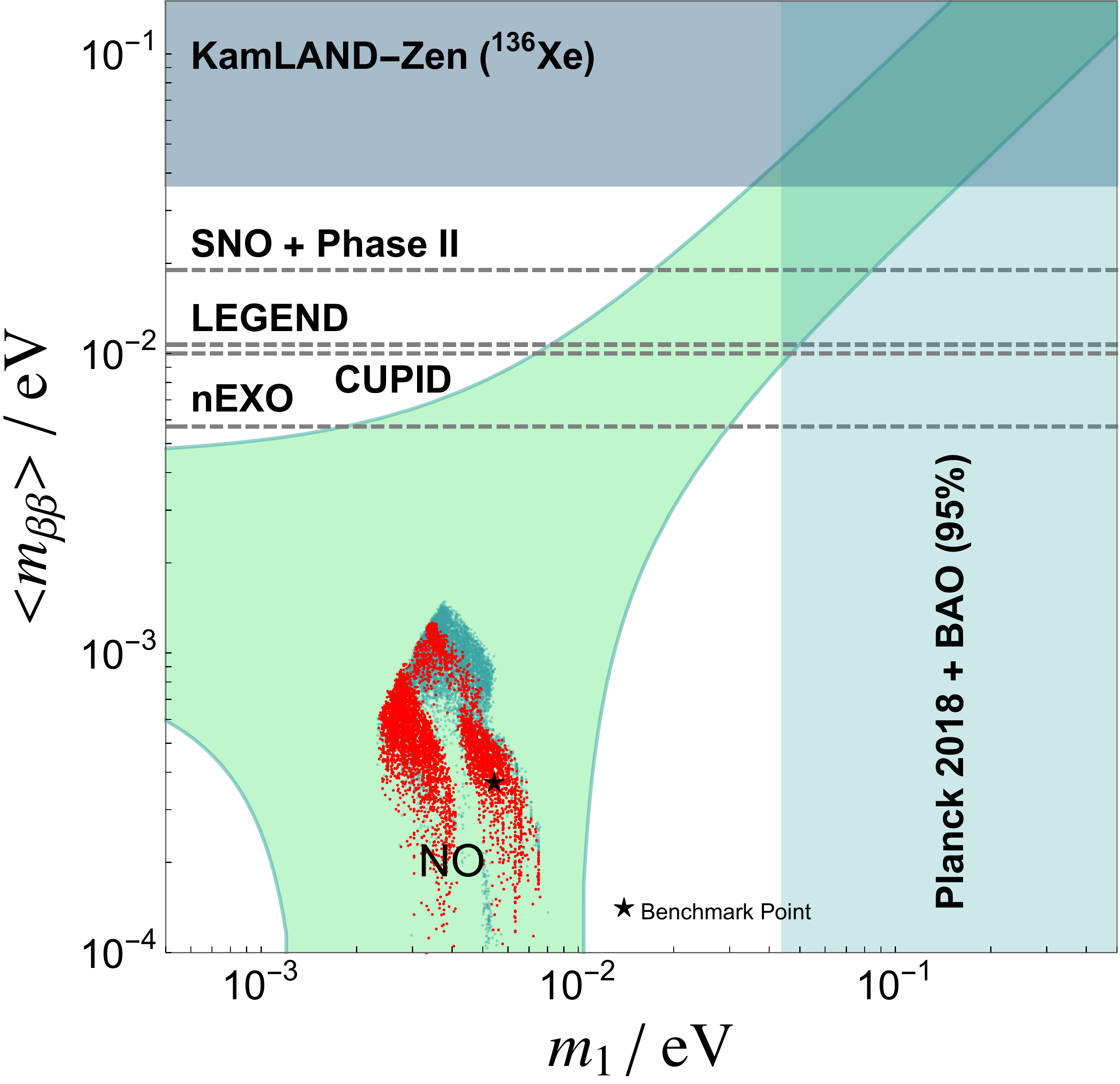}
    \caption{
    We display the viable points from this model in the $m^{\nu}_1$ vs. $\vev{m_{\beta \beta}}$. The region in green is the constraint coming from the mixing parameters at 3$\sigma$ \cite{deSalas:2020pgw}. We include experimental bounds on $\vev{m_{\beta \beta}}$; and $\sum m_{\nu}$ from cosmological observations \cite{KamLAND-Zen:2022tow,Planck:2018vyg}, also some projected sensitivities; LEGEND \cite{LEGEND:2017cdu},
 SNO + Phase II \cite{SNO:2015wyx}, nEXO \cite{nEXO:2017nam}, and CUPID \cite{CUPID:2019imh} shown as horizontal dashed lines.
    }
\label{fig:NuLess}
\end{figure}

Furthermore, since neutrinos are Majorana particles in this model, Lepton number violating (LNV) processes are present, such as neutrinoless double-beta decay ($\nu0\beta\beta$). In this case, a convenient description of the lepton mixing matrix is the symmetrical parameterization proposed in~\cite{Schechter:1980gr} and revisited in~\cite{Rodejohann:2011vc}.
The three physical phases are parameterized by $\phi_{12}$, $\phi_{13}$, and $\phi_{23}$, the leptonic CP-phase is given by
\begin{equation}
\delta^{\ell} = \phi_{13} - \phi_{12} -\phi_{23}\,,
\label{eq:CPPhaseSymm}
\end{equation}
while the phases $\phi_{12}$, and $\phi_{13}$ are crucial to describe lepton number violating processes. 
The amplitude associated with neutrinoless double-beta decay can be written in terms of the lepton mixing parameters as follows 
\begin{equation}
\vev{m_{\beta \beta}} = \left\lvert \cos^2\theta^{\ell}_{12} \cos^2\theta^{\ell}_{13} m^{\nu}_1 + \sin^2\theta^{\ell}_{12} \cos^2\theta^{\ell}_{13} m^{\nu}_2 e^{2 i \phi_{12}} +  \sin^2\theta^{\ell}_{13}  m^{\nu}_3 e^{2 i \phi_{13}}\right\rvert.
\end{equation}
In Figure \ref{fig:NuLess}, we display the values of the lightest neutrino mass $m^{\nu}_1$ vs.\ the mass parameter $\vev{m_{\beta \beta}}$ in this model. The region in green is the one consistent with experimental values for oscillation parameters at 3$\sigma$ \cite{deSalas:2020pgw}, while the different two thick bands represent the current experimental bounds on $\vev{m_{\beta \beta}}$, and the bound on $\sum m_{\nu}$ from cosmology \cite{KamLAND-Zen:2022tow,Planck:2018vyg}. The projected sensitivities for three different neutrinoless double-beta decay experiments, LEGEND \cite{LEGEND:2017cdu},
 SNO + Phase II \cite{SNO:2015wyx}, nEXO \cite{nEXO:2017nam}, and CUPID \cite{CUPID:2019imh} are displayed as horizontal dashed lines. 

\subsection{Dark Matter}

This work focuses on neutrino properties, and we restrict ourselves to a qualitative discussion of the resulting dark matter phenomenology.
The lightest $\mathbb{Z}_2$-odd particle is a dark matter candidate. Since for parameters consistent with lepton mixing, the heavy fermion mass $m_N$ is much larger than that of the neutral dark scalars in Eq.~(\ref{eq:BasisChange-Dark-Neutral}), we focus on the lightest dark scalar $\chi_4^D$ as WIMP dark matter candidate.

In the limit where the dark matter candidate is much lighter than the other dark scalars, the model resembles the real scalar singlet model (see eg.~\cite{Burgess:2000yq,Cline:2013gha}), which has only the DM mass $m_\chi$ and quartic coupling term $\frac{1}{2}\lambda_{\chi H} \chi^2 |H|^2$ as free parameters.
The dominant production process is annihilation through an $s$-channel Higgs into a pair of SM particles.
The most constraining signature is the spin-independent scattering of Galactic DM particles off nuclei, which likewise proceeds through exchanging a virtual Higgs boson.
Two mass ranges of scalar singlet DM are compatible with constraints: Either $m_\mathrm{DM} \lesssim m_h/2$, such that efficient annihilation through an on-shell Higgs allows for small values of the portal coupling $\lambda_{\chi H}$, suppressing DM direct detection (DMDD). Alternatively, the DM mass must be in the TeV range to avoid DMDD constraints.
In the present model, intermediate DM masses may also be viable, particularly for compressed dark scalar mass spectra that allow for efficient co-annihilation~\cite{Griest:1990kh}, or in the case of relatively light dark scalars that would enable for $t$-channel annihilation into pairs of EW gauge bosons.

\subsection{Benchmark Point}
\label{sec:benchmark}

\begin{table}[ht]
	\centering
	\scriptsize
	\renewcommand{\arraystretch}{.9}
	\begin{tabular}[t]{|l|c|}
		\hline
		Parameter & Value \\ 
		\hline
		$y_e$ & $3.96 \times 10^{-6}$ \\
		$y_{\mu}$ & $8.35 \times 10^{-4}$ \\
		$y_{\tau}$ &$ 1.42\times 10^{-2}$ \\
		$y^{\nu}_1$ & $-1.41\times 10^{-5}$ \\
		$y^{\nu}_2$ &$8.05\times 10^{-5}$ \\
		$y^{\nu}_3$ & $-1.47\times 10^{-4}$ \\
		$v_{\eta_1}/\mathrm{GeV}$ & $173.94$ \\
		$v_{H}/\mathrm{GeV}$ & $173.95$\\
		$m_N/\mathrm{GeV}$ & $9.59 \times 10^{6} $\\
		$\lambda_1$ & $0.732$ \\
		$\lambda_2$ & $3.5$ \\
		$\lambda_3$ & $-2.532$ \\
		$\lambda_4$ & $-1.205$ \\
		$\lambda_5$ & $1.16$ \\
		$\lambda_6$ & $3.492$ \\
		$\lambda_7$ & $3.489$ \\
		$\lambda_8$ & $-1.017$ \\
		$\lambda_9$ & $-1.118$ \\
		$\lambda_{10}$ & $-0.7$ \\
		$\varphi_{5}$ & $0.524$ \\
		$\varphi_{9}$ & $0.562$ \\
		$\varphi_{10}$ &$2.134$ \\
		\hline	
	\end{tabular}  
	\begin{tabular}[t]{ |l|c|c|c| }
		\hline
		\multirow{2}{*}{Observable}& \multicolumn{2}{c|}{Data} & \multirow{2}{*}{Best fit}  \\
		\cline{2-3}
		& Central value & 1$\sigma$ range   & \\
		\hline
		$\sin^2\theta_{12}/10^{-1}$ & 3.18 & 3.02 $\to$ 3.34 & $2.98$  \\ 
		$\sin^2\theta_{13}/10^{-2}$ (NO) & 2.200 & 2.138 $\to$ 2.269 & $2.222$  \\
		$\sin^2\theta_{23}/10^{-1}$ (NO) & 5.74 & 5.60 $\to$ 5.88 & $5.82$  \\
		$\delta^\ell$ $/\pi$ (NO) & 1.08 & 0.96 $\to$ 1.21 & $1.00$  \\
		$\Delta m_{21}^2 / (10^{-5} \, \mathrm{eV}^2 ) $  & 7.50  & 7.30 $\to$ 7.72 & $7.43$  \\
		$\Delta m_{31}^2 / (10^{-3} \, \mathrm{eV}^2) $ (NO) & 2.55  & 2.52 $\to$ 2.57 &  $2.55$ \\
		$m^{\nu}_{\text{lightest}}$ $/\mathrm{meV}$  (NO) & & & $ 5.45 $ \\ 
		$m^{\nu}_2$ $/\mathrm{meV}$  & &&  $ 10.20$ \\ 
		$m^{\nu}_3$ $/\mathrm{meV}$  & &&  $50.81 $ \\
		$ \phi_{12}/ \pi $ & & & $0.5$  \\
		$ \phi_{13}/ \pi$ & & & $0.5$  \\    
		$ \phi_{23}/ \pi$ & & & $1.0$  \\
		$ \langle m_{\beta \beta} \rangle/ \mathrm{eV}$ & & & $3.66\times 10^{-4}$  \\
		$m_e$ $/ \mathrm{MeV}$ & 0.486 &  0.486 $\to$ 0.486 & $0.486$ \\ 
		$m_\mu$ $/  \mathrm{GeV}$ & 0.102 & 0.102  $\to$ 0.102  &  $0.102$ \\ 
		$m_\tau$ $/ \mathrm{GeV}$ &1.746 & 1.743 $\to$1.747 & $1.746$ \\     	
		\hline
         $v/\mathrm{GeV}$ &246 & 246 $\to$ 246 &  $246$\\
		$M_H/\mathrm{GeV}$  (Higgs boson)&125.25 & 125.08 $\to$ 125.42 &  $125.30$ \\	
		$M_{DM}/\mathrm{GeV}$  (scalar DM)   & & & $59$  \\
		$M_N/\mathrm{GeV}$ & & & $9.59 \times 10^{6} $ \\	
		\hline
		$M_{H_{0}}/\mathrm{GeV}$  \text{(Heavy Higgs)}& &  &  $295$ \\	
		$M_{A_0}/\mathrm{GeV}$  \text{(Pseudoscalar)}   & & & $701$  \\
		$M^{+}_{H_0}/\mathrm{GeV}$ \text{(Active)} & & & $375$ \\	
		$M_{\chi^+_1}/\mathrm{GeV}$ \text{(Dark)} & & & $469$ \\	
		$M_{\chi^+_2}/\mathrm{GeV}$ \text{(Dark)} & & & $388$ \\
		$M^{0}_{\chi_1}/\mathrm{GeV}$ \text{(Dark)} & & & $618$ \\
		$M^{0}_{\chi_2}/\mathrm{GeV}$ \text{(Dark)} & & & $547$ \\
		$M^{0}_{\chi_3}/\mathrm{GeV}$ \text{(Dark)} & & & $440$ \\
        $\lambda_{\chi h}$ & & & $1.0\times 10^{-3}$ \\
		\hline
	\end{tabular}
        \caption{
    Model parameters and corresponding observables: measured values compared to values for this benchmark point.} 
	\label{tab:fit}
\end{table}

 We summarize the results from a selected benchmark point from our parameter space scan in Table \ref{tab:fit}. It includes the numerical values of the model's parameters and the corresponding values of various observables, the oscillation parameters, the masses of the charged leptons and neutrinos, and the scalar mass spectrum. For this benchmark point, all the listed observables are within their experimental 1$\sigma$ region.
 In addition, we verified that the benchmark point is marginally compatible with Higgs signal strength constraints~\cite{ATLAS:2021vrm,Atkinson:2022pcn}, as well as the measured Higgs branching fractions to photons and invisible final states~\cite{Gunion:1989we,ATLAS:2022tnm,CMS:2022qva}.
 This benchmark point is marked with a star in Figures \ref{fig:AtmVsCP}, and \ref{fig:NuLess}.

The dark matter phenomenology at the benchmark point is governed by the term $\frac{1}{2}\lambda_{\chi h} \chi_4^0 \chi_4^0 h$ in the broken-phase scalar potential. The value of $\lambda_{\chi h}$ depends on a combination of quartic couplings and dark scalar mixing angles and is listed separately in Table~\ref{tab:fit}. This directly maps to the real scalar singlet model in terms of DM direct detection. Comparing to previous studies~\cite{Cline:2013gha}, it is clear that the benchmark is compatible with current DMDD experiments.
The comparison with the scalar singlet model also shows that the benchmark point is compatible with DM overabundance constraints, even before considering the additional annihilation channels possible in the present scenario.

\section{Discussion and conclusions}
The difference in scale between the solar and atmospheric mass splitting, ($\lvert \Delta m^{2}_{31} \rvert \gg \Delta m^{2}_{21}$), was explained in previous works with a tree-level seesaw and a one-loop scotogenic mechanism. In this work, we have studied a model based on a $A_4$ flavor symmetry in the lepton sector, the Discrete Dark Matter (DDM) model, in which the interplay between the seesaw and scotogenic mechanisms arises directly from a remnant symmetry after the spontaneous $A_4$ breaking. This residual symmetry stabilizes a scalar dark matter candidate.
The model has the unique property among the scoto-seesaw models that the Yukawa couplings ruling both mechanisms are the same. Therefore tree-level seesaw dominance over the one-loop scotogenic mechanism is not imposed, but it is a prediction of the model as a consequence of the $A_4$ flavor symmetry. 

Furthermore, we would like to remark compelling features of the model:
\begin{itemize}
\item It predicts normal ordering (NO) for neutrino masses, favored by neutrino oscillation experiments.
\item Due to the flavor symmetry, the only source of leptonic CP-violation originates in the scalar potential. Furthermore, the latter is necessary to obtain lepton mixing parameters consistent with the current experimental values. The values obtained for the Dirac CP-phase lie close to the CP-preserving values, $\delta^{\ell} \sim 0,\pi, 2\pi$, as displayed in Figure \ref{fig:AtmVsCP}.
\item There is a strong correlation between the mass of the lightest neutrino $m^{\nu}_1$ (NO) and the solar mixing angle $\theta_{12}$, which constrains $m^{\nu}_1$ to be between $2\hspace{2pt} \text{meV}\hspace{2pt}\lesssim \hspace{2pt} m^{\nu}_{\text{lightest}} \hspace{2pt} \lesssim \hspace{2pt} 8 \hspace{2pt} \text{meV}$.
\end{itemize}

The model contains a rich scalar sector with charged and neutral states, accessible to collider experiments. As we mentioned, the active sector is similar to the type-I 2HDM. Therefore, the production cross section at NNLO level and dominant decay channels are similar to those given in Ref. \cite{Chen:2019pkq}. On the other hand, the dark scalars are characterized of being produced in pairs due to the $\mathbb{Z}_2$ residual symmetry.
These dark pairs can be produced via vector boson fusion or the Drell-Yan mechanism as described in \cite{Dutta:2017lny}.
A detailed phenomenological study of the scalar sector is beyond the scope of this paper. A previous work on the DDM model explores this sector \cite{Hirsch:2010ru,Boucenna:2011tj}.

\begin{acknowledgments} 
Work supported by the Chilean grant Fondecyt Iniciaci\'on No. 11201240 (Nu Physics); the Mexican grants CONACYT CB-2017-2018/A1-S-13051 and DGAPA-PAPIIT IN107621; and the Spanish grants PID2020-113775GB-I00 (AEI/10.13039/501100011033) and Prometeo CIPROM/2021/054 (Generalitat Valenciana). OM is supported by  Programa Santiago Grisolía (No. GRISOLIA/2020/025). EP is grateful to funding from `C\'atedras Marcos Moshinsky' (Fundaci\'on Marcos Moshinsky). OM thanks Salvador Centelles Chuliá, Professor Luís Lavoura for insightful discussion through email correspondence, and Professor Martin Konrad Hirsch for his ($\infty$) patience and help during discussions concerning this work. 
CB would like to thank IFUNAM and Instituto de Física
Corpuscular (CSIC-UV) for the hospitality while part of this work was carried out. 

\end{acknowledgments}

\appendix
\section{The $A_4$ Basis and its Representation Products}
\label{Appendix:A4}

The $A_4$ group can be defined by its presentation equation
\begin{equation}
A_4\simeq \{S,T|S^2=T^3=(ST)^2=1\}.
\end{equation}
With two generators $S$, and $T$. The $A_4$ group has four irreducible representations: three singlets $\bf{1}$, $\bf{1^{\prime}}$, and $\bf{1^{\prime\prime}}$ and one triplet $\bf{3}$. On the Ma-Rajasekaran basis, the generators are
\begin{equation}\begin{split}
\textbf{1}&:\ \ \ \ S=1,\ \ \ \ T=1,\\
\textbf{1}'&:\ \ \ \ S=1,\ \ \ \ T=\omega,\\
\textbf{1}''&:\ \ \ \ S=1,\ \ \ \ T=\omega^2,\\
\textbf{3}&:\ \ \ \ S=\left(\begin{array}{ccc} 1&0&0\\ 0&-1&0\\ 0&0&-1
\end{array}\right),\ \ \ \ T=\left(\begin{array}{ccc} 0&1&0\\ 0&0&1\\ 1&0&0
\end{array}\right),\\
\end{split}\label{eq:A4Basis}\end{equation}
where $\omega=e^{2i\pi/3}$. Notice that in this basis the $S$ generator of the triplet representation is diagonal. 

The $A_4$ representations have, independently of the basis, the following non-trivial contractions
\begin{equation}\begin{split}
\textbf{1}'\otimes\textbf{1}'=&\textbf{1}'',\ \ \ \textbf{1}''\otimes\textbf{1}''=\textbf{1}',\ \ \ \textbf{1}''\otimes\textbf{1}'=\textbf{1},\\
\textbf{3}&\otimes \textbf{3} =\textbf{1}\oplus\textbf{1}'\oplus\textbf{1}''\oplus\textbf{3}_1\oplus\textbf{3}_2.
\end{split}\end{equation}
The contractions of two triplets $\textbf{3}_a\sim (a_1,a_2,a_3)$ and $\textbf{3}_b\sim (b_1,b_2,b_3)$, in the chosen basis, are decomposed as
\begin{equation}\begin{split}
\textbf{3}_a\otimes \textbf{3}_b\to \textbf{1}&=a_1b_1+a_2b_2+a_3b_3,\\\textbf{3}_a\otimes \textbf{3}_b\to \textbf{1}'&=a_1b_1+\omega^2 a_2b_2+\omega a_3b_3,\\
\textbf{3}_a\otimes \textbf{3}_b\to \textbf{1}''&=a_1b_1+\omega a_2b_2+\omega^2 a_3b_3,\\
\textbf{3}_a\otimes \textbf{3}_b\to \textbf{3}_1&=(a_2b_3,a_3b_1,a_1b_2),\\
\textbf{3}_a\otimes \textbf{3}_b\to \textbf{3}_2&=(a_3b_2,a_1b_3,a_2b_1).\\
\end{split}\end{equation}

\section{The Scalar Potential}
\label{Appendix:scalarpot}
Given the field content in Table \ref{tab:Mod} the most general scalar potential invariant under the $SU(2)_L\times U(1)_Y\times A_4$ symmetry is
{\small\begin{align}
V(H,\eta)&= \mu^2_H H^{\dagger}H + \mu^2_{\eta} \left(\eta^{\dagger} \eta \right)_{\mathbf{1}} +  \lambda_1 \left(H^{\dagger} H \right)^2 + \lambda_2 \left( \eta^{\dagger} \eta \right)_{\mathbf{1}} \left( \eta^{\dagger} \eta  \right)_{\mathbf{1}} + \lambda_3 \left( \eta^{\dagger} \eta \right)_{\mathbf{1'}} \left( \eta^{\dagger} \eta  \right)_{\mathbf{1''}} +  \lambda_4 \left( \eta^{\dagger} \eta \right)_{\mathbf{3}_1} \left( \eta^{\dagger} \eta  \right)_{\mathbf{3}_2} \nonumber \\
&\hspace{2pt}  +  \left[ \lambda_5 e^{i \varphi_5} \left( \eta^{\dagger} \eta \right)_{\mathbf{3}_1} \left( \eta^{\dagger} \eta  \right)_{\mathbf{3}_1}  + \textit{h.c.}\right] + \lambda_6 \left( H^{\dagger} H \right) \left(\eta^{\dagger} \eta \right)_{\mathbf{1}} + \lambda_7 \left( H^{\dagger} \eta \right) \left(\eta^{\dagger} H \right)  + \left[ \lambda_8 e^{i \varphi_8} \left( H^{\dagger} \eta \right) \left( H^{\dagger} \eta  \right)  + \textit{h.c.}\right] \nonumber \\
&\hspace{2pt}+  \left[ \lambda_9 e^{i \varphi_9} \left( \eta^{\dagger} \eta \right)_{\mathbf{3}_1} \left( H^{\dagger} \eta  \right)  + \textit{h.c.}\right] +  \left[ \lambda_{10} e^{i \varphi_{10}} \left( \eta^{\dagger} \eta \right)_{\mathbf{3}_2} \left( H^{\dagger} \eta  \right)  + \textit{h.c.}\right]
\label{eq:ScalarPotential}
\end{align}}
There are four potentially complex parameters in the scalar potential, and we chose to parameterize it in terms of sixteen real parameters:
\begin{equation}
\{ \mu^2_H,\hspace{2pt} \mu^2_{\eta},\hspace{2pt} \lambda_1,\hspace{2pt}\lambda_2,\hspace{2pt} \lambda_3,\hspace{2pt} \lambda_4,\hspace{2pt} \lambda_5,\hspace{2pt} \varphi_5,\hspace{2pt} \lambda_6,\hspace{2pt} \lambda_7,\hspace{2pt}\lambda_8,\hspace{2pt} \varphi_8,\hspace{2pt}\lambda_9,\hspace{2pt} \varphi_9,\hspace{2pt}\lambda_{10},\hspace{2pt} \varphi_{10}\}.
\label{eq:ScalarPotentialParam}
\end{equation}
Furthermore, we set to zero the phase with argument $\varphi_8$ by global re-phasing of the fields. Hence it does not appear in our results as a physical parameter of the model.
The minimization conditions are given by
\begin{align}
\mu^2_{H} &=\frac{1}{2} \left(-2 v_{s}^{2} \lambda_{1}-v_{1}^{2} \lambda_{6}-v_{1}^{2} \lambda_{7}-8 v_{1}^{2} \lambda_{8} \right), \label{eq:muH} \\
\mu^2_{\eta} &=\frac{1}{2} \left(-2 v_{1}^{2} \lambda_{2}-2 v_{1}^{2} \lambda_{3}-v_{s}^{2} \lambda_{6}-v_{s}^{2} \lambda_{7}-8 v_{s}^{2} \lambda_{8} \right).\label{eq:mueta}
\end{align}
After electroweak and flavor symmetry breakdown, as described by Eq. (\ref{eq:Vevs}), we obtain the mass matrix for the active neutral scalars
\begin{equation}
M^2_{H_0^{\prime} H_1^{\prime}}= \begin{pmatrix}  \vess (2 \lambda_1)& \vone \ves \left( \lambda_6 + \lambda_7 + 8\lambda_8 \right)  \\
\vone \ves \left( \lambda_6 + \lambda_7 + 8\lambda_8 \right) &  \vones \left(2 \lambda_2 + 2\lambda_3 \right)\\ \end{pmatrix}, 
\hspace{2pt} 
M^2_{A_0^{\prime}A_1^{\prime}}= \begin{pmatrix}-\vones \left(8\lambda_8\right) & \vone \ves \left(8\lambda_8\right) \\
\vone \ves \left(8\lambda_8\right) & -\vess \left(8\lambda_8\right)\\\end{pmatrix}.
\label{eq:M-ActiveNeutral}
\end{equation}
As mentioned in previous sections, it is necessary to violate CP in the dark scalar sector to obtain the correct neutrino oscillation parameters in this model. Following the notation defined by Eq. (\ref{eq:blockdiag}) we display separately the mass matrix of dark scalars $M^2_{H_2^{\prime}H_3^{\prime}}$, dark pseudo-scalars $M^2_{A_2^{\prime}A_3^{\prime}}$ and the matrix that generates a mixing between these fields $M^2_{\text{CPV}}$, explicitly violating CP symmetry
\begin{equation}
M^2_{H_2^{\prime}H_3^{\prime}}= \begin{pmatrix} \vones\left(-\frac{3}{2}\lambda_3+\frac{1}{2}\lambda_4 +\lambda_5\cos{\varphi_5}\right)&6\vone\ves\left(  \lambda_9\cos{\varphi_{9}} + \lambda_{10}\cos{\varphi_{10}}\right)  \\
6\vone\ves\left( \lambda_9\cos{\varphi_{9}} + \lambda_{10}\cos{\varphi_{10}} \right)&   \vones\left(-\frac{3}{2}\lambda_3+\frac{1}{2}\lambda_4 +\lambda_5\cos{\varphi_5}\right)\\ \end{pmatrix},
\label{eq:M-DarkNeutralS}
\end{equation}

\begin{equation}
M^2_{A_2^{\prime}A_3^{\prime}}= \begin{pmatrix} \vones \left( - \frac{3}{2} \lambda_3 + \frac{1}{2} \lambda_4 -  \lambda_5 \cos{\varphi_5} \right) - \vess \left( 8 \lambda_8 \right) & 2 \vone \ves \left( \lambda_9 \cos{\varphi_9} +  \lambda_{10} \cos{\varphi_{10}} \right) \\
 2 \vone \ves \left( \lambda_9 \cos{\varphi_9} +  \lambda_{10} \cos{\varphi_{10}} \right) & \vones \left( - \frac{3}{2} \lambda_3 + \frac{1}{2} \lambda_4 -  \lambda_5 \cos{\varphi_5} \right) - \vess \left( 8 \lambda_8 \right)\\\end{pmatrix}, 
 \label{eq:M-ActiveNeutralP}
\end{equation}

\begin{equation}
M^2_{\text{CPV}}= \begin{pmatrix} - \vones \left( \lambda_5 \sin{\varphi_5}\right) &-2  \vone \ves \left( \lambda_9 \sin{\varphi_9}  + \lambda_{10} \sin{\varphi_{10}} \right) \\
-2  \vone \ves \left( \lambda_9 \sin{\varphi_9}  + \lambda_{10} \sin{\varphi_{10}} \right) & \vones \left( \lambda_5 \sin{\varphi_5}\right)
  \end{pmatrix}.
  \label{eq:M-ActiveNeutralCP}
\end{equation}
The mixing matrix for the active charged scalars is given by
\begin{equation}
M^2_{H_0^{\prime+}H_1^{\prime+}}= \begin{pmatrix} - \vones \left( \frac{1}{2} \lambda_7+ 4 \lambda_8\right) & \vone \ves \left( \frac{1}{2} \lambda_7+ 4 \lambda_8\right) \\
\vone \ves \left( \frac{1}{2} \lambda_7+ 4 \lambda_8\right) & -\vess \left( \frac{1}{2} \lambda_7+ 4 \lambda_8\right)
  \end{pmatrix},
  \label{eq:M-ActiveCharged}
\end{equation}
and lastly the mass matrix of the dark charged scalars
\begin{equation}
M^2_{H_2^{\prime+}H_3^{\prime+}} = \begin{pmatrix} - \vones \left( \frac{3}{2} \lambda_3\right) - \vess \left( \frac{1}{2} \lambda_7 + 4 \lambda_8 \right) &  2 \vone \ves \left( \lambda_{9} e^{-i \varphi_{9}}+ \lambda_{10} e^{i \varphi_{10}}\right) \\
 2 \vone \ves \left( \lambda_{9} e^{i \varphi_{9}}+ \lambda_{10} e^{-i \varphi_{10}}\right) & - \vones \left( \frac{3}{2} \lambda_3\right) - \vess \left( \frac{1}{2} \lambda_7 + 4 \lambda_8 \right)
  \end{pmatrix}.
  \label{eq:M-DarkCharged}
\end{equation}
We were able to write the expressions for the spectrum of mass-eigenstates in the active sector, in which CP is conserved.
\begin{eqnarray}
 \label{mass_eq}
 M^2_{H_0}&=&\lambda_1 v^2_H +P v^2_{\eta_1} +\sqrt {v^2_{\eta_1} v^2_H (L^2-2 P \lambda_1)+(P v^2_{\eta_1} +\lambda_1 v^2_H
  )^2 },\\
  M_{A_0}^2&=&-8 v^2 \lambda_{8}, \\
  M_{h}^2&=&\lambda_1 v^2_H +P v^2_{\eta_1} -\sqrt{ v^2_{\eta_1} v^2_H (L^2-2 P \lambda_1)+(P v^2_{\eta_1} +\lambda_1 v^2_H )^2}
  , \\
  M^2_{H^+_{0}}&=&-(\lambda_{7}+8\lambda_{8}) v^2/2,
\end{eqnarray}
where to simplify the notation for the scalar masses, we defined the following variables:
\begin{align}
L&=\lambda_6+\lambda_{7}+8\lambda_{8}, \\
P&=\lambda_2 +\lambda_3.
\end{align}

\section{Other Scenarios}
\label{Appendix:OtherScenarios}
In the results presented in the previous sections, we considered the $A_4$ representation assignments for the lepton doublets as in Table \ref{tab:Mod}:
\begin{equation}
\label{eq:Scenario1}
L_{e} \sim \mathbf{1}^{\prime \prime}, \qquad  L_{\mu} \sim \mathbf{1}, \qquad L_{\tau} \sim \mathbf{1}^{\prime}.
\end{equation}
Nonetheless, we analyzed the parameter space of the other configurations of the charge assignments that keep a diagonal charge lepton mass matrix, as in Eq. (\ref{eq:MDirA}). The other two independent possibilities are 
\begin{equation}
\label{eq:Scenario1}
L_{e} \sim \mathbf{1}, \qquad  L_{\mu} \sim \mathbf{1}^{\prime}, \qquad L_{\tau} \sim \mathbf{1}^{\prime \prime}, \qquad \text{and} \qquad 
L_{e} \sim \mathbf{1}^{\prime}, \qquad  L_{\mu} \sim \mathbf{1}^{\prime \prime}, \qquad L_{\tau} \sim \mathbf{1}.
\end{equation}
We find that in this model, the three scenarios result in the same neutrino phenomenology, ie.\ the results displayed in Figures \ref{fig:AtmVsCP}, \ref{fig:SolVsm1}, and \ref{fig:NuLess} are independent of the $A_4$ charge assignments of the Lepton doublets.

As was argued above, we also analyzed the scenario of CP preservation in the scalar potential by taking
\begin{equation}
\varphi_5,\hspace{2pt}\varphi_9,\hspace{2pt}\varphi_{10}=0.
\label{eq:VanishingPhases}
\end{equation}
which simplifies the analysis of neutrino mass generation outlined in Section \ref{Section:NeutrinoMasses} because the matrix $M^{2}_{\text{CPV}}$ in Eqs. (\ref{eq:M-ActiveNeutralCP}), and (\ref{eq:RotMat}) vanishes. In such a case, the mixing angle between the dark scalars $\eta_2$ and $\eta_3$ is $\pi/4$. This scenario is ruled out since it is impossible to fit all the oscillation parameters within their allowed 3$\sigma$ range. More specifically, the predicted solar mixing angle $\sin^2{\theta_{12}}$ is very small compared to its central value \cite{deSalas:2020pgw}, the best-fit point found for this scenario is summarized in Table \ref{tab:fitNOCP}.
\begin{table}[htb]
	\centering
	\scriptsize
	\renewcommand{\arraystretch}{.9}
	\begin{tabular}[t]{|l|r|}
		\hline
		Parameter & Value \\ 
		\hline
		$y_e$ &$3.34\times 10^{-6}$ \\
		$y_{\mu}$ & $7.06 \times 10^{-4}$ \\
		$y_{\tau}$ &$ 1.12\times 10^{-2}$ \\
		$y^{\nu}_1$ & $-1.16\times 10^{-5}$ \\
		$y^{\nu}_2$ & $5.69\times 10^{-5}$ \\
		$y^{\nu}_3$ & $-6.37\times 10^{-5}$ \\
		$v_{\eta_1}/\mathrm{GeV}$ & $134.91$ \\
		$v_{H}/\mathrm{GeV}$ & $205.71$\\
		$M_N/\mathrm{GeV}$ & $1.62\times 10^{6}$\\
		$\lambda_1$ & $0.38$ \\
		$\lambda_2$ & $3.17$ \\
		$\lambda_3$ & $-1.11$ \\
		$\lambda_4$ & $3.14$ \\
		$\lambda_5$ & $2.45$ \\
		$\lambda_6$ & $3.30$ \\
		$\lambda_7$ & $-1.16$ \\
		$\lambda_8$ & $-0.41$ \\
		$\lambda_9$ & $-1.94$ \\
		$\lambda_{10}$ & $2.54$ \\
		$\varphi_{5}$ & $0$ \\
		$\varphi_{9}$ & $0$ \\
		$\varphi_{10}$ & $0$ \\
		\hline	
	\end{tabular}  
	\begin{tabular}[t]{ |l |c|c|c| }
		\hline
		\multirow{2}{*}{Observable}& \multicolumn{2}{c|}{Data} & \multirow{2}{*}{Best fit} \\
		\cline{2-3}
		& Central value & 1$\sigma$ range     & \\
		\hline
		$\sin^2\theta_{12}/10^{-1}$ & 3.18 & 3.02 $\to$ 3.34 & $1.46$  \\ 
		$\sin^2\theta_{13}/10^{-2}$ (NO) & 2.200 & 2.138 $\to$ 2.269 & $2.314$  \\  
		$\sin^2\theta_{23}/10^{-1}$ (NO) & 5.74 & 5.60 $\to$ 5.88 & $6.2$  \\
		$\delta^\ell$ $/\pi$ (NO) & 1.08 & 0.96 $\to$ 1.21 & $1.0$  \\
		$\Delta m_{21}^2 / (10^{-5} \, \mathrm{eV}^2 ) $  & 7.50  & 7.30 $\to$ 7.72 & $7.49$  \\
		$\Delta m_{31}^2 / (10^{-3} \, \mathrm{eV}^2) $ (NO) & 2.55  & 2.52 $\to$ 2.57 &  $2.55$ \\
		$m^{\nu}_{\text{lightest}}$ $/\mathrm{meV}$  (NO) & & &  $ 2.03 $ \\ 
		$m^{\nu}_2$ $/\mathrm{meV}$  & &&  $ 8.89$ \\ 
		$m^{\nu}_3$ $/\mathrm{meV}$  & &&  $50.54 $ \\
		$ \phi_{12}/ \pi $ & & & $1.5$  \\
		$ \phi_{13}/ \pi$ & & & $1.5$  \\    
		$ \phi_{23}/ \pi$ & & & $1.0$  \\
		$ \langle m_{\beta \beta} \rangle/ \mathrm{eV}$ & & & $7.48\times 10^{-4}$  \\
		$m_e$ $/ \mathrm{MeV}$ & 0.486 &  0.486 $\to$ 0.486 & $0.486$ \\ 
		$m_\mu$ $/  \mathrm{GeV}$ & 0.102 & 0.102  $\to$ 0.102  &  $0.102$ \\ 
		$m_\tau$ $/ \mathrm{GeV}$ &1.746 & 1.743 $\to$1.747 & $1.746$ \\     	
		\hline
		$M_H/\mathrm{GeV}$  (Higgs boson)&125.25 & 125.08 $\to$ 125.42 &  $125.25$ \\	
		$M_{DM}/\mathrm{GeV}$  (Scalar DM)   & & & $64.95$  \\
		$M_N/\mathrm{GeV}$ & & & $1.62\times 10^{6}$ \\	
		\hline		
	\end{tabular}
        \caption{
        Model parameters and observables in the model with no CP violation in the scalar potential: measured values compared to values for the best fit point. Most of the observables obtained are inside the experimental 2$\sigma$ range, with the exception of the solar mixing angle $\sin^2\theta_{12}$, which is completely off the experimental central value. This makes this scenario unfeasible. } 
	\label{tab:fitNOCP}
\end{table}

\bibliographystyle{JHEP}
\bibliography{DarkA4-v1Notes.bib}

\end{document}